\documentclass[fleqn,10pt]{wlscirep}
\usepackage[utf8]{inputenc}
\usepackage[T1]{fontenc}
\usepackage{bm}

\title{Super-resolution photoacoustic and ultrasound imaging with sparse arrays}

\newcommand{\ignore}[1]{}
\newcommand{\bs}[1]{\boldsymbol#1}
\newcommand\rr{\mathbf{r}} 

\author[1]{Sergey Vilov}
\author[1]{Bastien Arnal}
\author[2]{Eliel Hojman}
\author[3]{Yonina C. Eldar}
\author[2]{Ori Katz}
\author[1,*]{Emmanuel Bossy}

\affil[1]{Univ. Grenoble Alpes, CNRS, LIPhy, 38000 Grenoble, France}
\affil[2]{Department of Applied Physics, Hebrew University of Jerusalem,  9190401 Jerusalem, Israel}
\affil[3]{Faculty of Mathematics and Computer Science, Weizmann Institute of Science, Rehovot, Israel}

\affil[*]{emmanuel.bossy@univ-grenoble-alpes.fr}

\begin{abstract}
It has previously been demonstrated that model-based reconstruction methods relying on \textit{a priori} knowledge of the imaging point spread function (PSF) coupled to sparsity priors on the object to image can provide super-resolution in photoacoustic (PA) or in ultrasound (US) imaging. Here, we experimentally show that such reconstruction also leads to super-resolution in both PA and US imaging with arrays having much less elements than  used conventionally (sparse arrays). As a proof of concept, we obtained super-resolution PA and US cross-sectional images of microfluidic channels with only 8 elements of a 128-elements linear array using a reconstruction approach based on a linear propagation forward model and assuming sparsity of the imaged structure.   Although the microchannels appear indistinguishable in the conventional delay-and-sum images obtained with all the 128 transducer elements, the applied sparsity-constrained model-based reconstruction provides super-resolution with down to only 8 elements. We also report simulation results showing that the minimal number of transducer elements required to obtain a correct reconstruction is fundamentally limited by the signal-to-noise ratio. The proposed method can be straigthforwardly applied to any transducer geometry, including 2D sparse arrays for 3D super-resolution PA and US imaging.
\end{abstract}
\begin{document}

\flushbottom
\maketitle
\thispagestyle{empty}

\section*{Introduction}

Ultrasound\cite{wells2006ultrasound} (US) and photoacoustic\cite{beard2011biomedical} (PA) imaging are now widely applied biomedical imaging modalities. They both usually use multielement transducer arrays as ultrasonic detectors for acquiring acoustic signals. Developed for two-dimensional (2D) or cross-sectional imaging, linear transducer arrays are widely spread in research and clinical applications. For three-dimensional (3D) single shot imaging, 2D array matrices should be used instead of linear arrays. However, the availability of 3D imaging equipment is first limited by the sophisticated fabrication process involved: probes for 3D imaging may have several thousands of elements to connect which makes the assembly of such probes technically difficult\cite{Fenster_2001}. Second, to control simultaneously as many elements as possible, sophisticated ultrasound electronics with a very large number of channels are needed. 
Decades ago, sparse arrays were proposed\cite{turnbull1991beam,smith1991high} to reduce substantially the number of transducer elements required for 3D US imaging. The corresponding reduction scheme suggests using a selection of elements of a dense periodic array without changing the total transducer aperture. Such a selection can either be a random subset of elements\cite{turnbull1991beam} or a defined pattern\cite{smith1991high}. Experimental investigations of imaging performance of sparse arrays in US imaging showed that the sparse array can provide a diffraction-limited resolution similar  \cite{roux20173d, roux2017validation} to that of the all-element array.  Special reconstruction techniques have been recently proposed for sparse array US imaging.  In particular, the convolutional beamforming algorithm was reported\cite{cohen2018sparse} to provide better (although still diffraction-limited) resolution than standard delay-and-sum beamforming. It is worth noting that in radar technologies it has also been shown that advanced compressed sensing methods\cite{eldar2012compressed} permit preserving the diffraction -limited resolution when using radars with a reduced number of elements\cite{rossi2013spatial,cohen2018summer}.

Sparse arrays have also been widely applied in 3D photoacoustic tomography\cite{ephrat2008three,chaudhary2010comparison,roumeliotis2011singular,roumeliotis20123d,wang2012investigation,dean2012accurate,han2017three}. In most of these studies  sparse arrays were used to obtain diffraction-limited images with model-based reconstruction\cite{chaudhary2010comparison,roumeliotis2011singular,roumeliotis20123d,wang2012investigation,dean2012accurate,han2017three}. The model-based reconstruction approach considers a forward linear model expressed as $S=\bs{A}T$, where $S$ are the acquired signals, $T$ is the object to reconstruct, and the propagation matrix $\bs{A}$ is a library containing the point spread function (PSF) at all points of the discretized imaging zone. The reconstruction consists in finding the object that minimizes a cost function defined as the sum of a fidelity term (taking into account the measurement data and the model) and a regularization term (taking into account measurement noise and prior knowledge on the object). 

In this work, we investigate the possibility of using sparse ultrasound arrays in the context of \textit{super-resolution} imaging with the model-based approach. The classical resolution in both PA and US imaging of biological tissues at depth is limited by acoustic diffraction. Specifically, it is the absorption of ultrasound in tissues that limits the highest ultrasound frequency detectable at a given depth, and therefore the resolution via the diffraction phenomenon. In particular, the depth-to-resolution ratio turns out to be of the order of 100 for both US and PA imaging of biological tissues. As an example, for a 10 MHz US probe, that can be ordinarily used for US or PA imaging at a few centimeters depth in tissues, diffraction limits the resolution to 100-200 $\mu$m  in typical imaging conditions\cite{foster2000advances}. 

Imaging beyond the diffraction limit has first been investigated in optical microscopy,  resulting in the advent of Nobel-prize winning pioneering methods such as photo-activated localization  microscopy (PALM)\cite{betzig2006imaging} and stimulated emission depletion (STED) microscopy\cite{hell1994breaking}.
Following the aforementioned advances in optics, several super-resolution techniques have been proposed more recently for both US and PA imaging.
Super-resolution methods include localization-based imaging, fluctuation-based reconstruction and model-based reconstruction. 
As for optical imaging, the localization approach in US\cite{viessmann2013acoustic,oreilly2013super,desailly2013sono} and PA\cite{vilov2017overcoming,dean2018localization} imaging relies on detection of individual scatterers or absorbers. The idea is that the coordinates of a point-like source can be determined with a precision much better than the size of the imaging PSF provided that this PSF can be separated from those of the other sources in some parameter space. This separation condition imposes a low concentration of sources, and therefore the use of contrast agents for visualization of biological structures. Fluctuation-based PA\cite{chaigne2016super,chaigne2017super} and US\cite{bar2017} techniques, based on the principles of super-resolution optical fluctuation imaging (SOFI)  exploit uncorrelated fluctuations from different sources. While fluctuation-based approaches eliminate the need for isolating individual sources, they remain limited in terms of spatial resolution improvement and temporal resolution.
It was also shown that model-based reconstruction with sparsity constraints on the sample, an approach originating from the field of compressed sensing\cite{eldar2012compressed}, could yield super-resolved  US~\cite{tuysuzoglu2012sparsity,zhao2016single} and PA\cite{egolf2018sparsity} images. A major advantage of model-based reconstruction over the previously mentioned localization-based and fluctuation-based techniques is that by requiring (in principle) only a single-shot acquisition it permits a high temporal resolution. Some very recent works~\cite{hojman2017photoacoustic, murray2017super, bar2018sparsity} have proposed to mix fluctuation-based and model-based techniques to achieve super-resolution in US~\cite{bar2018sparsity} and PA imaging~\cite{murray2017super}. In PA imaging, it was also demonstrated\cite{hojman2017photoacoustic} that super-resolution can be obtained via the joint support recovery through the model-based approach. In particular, the vector $S$ is composed of data measured at several PA acquisitions, with each acquisition corresponding to a random speckle illumination of \emph{the same} absorbing structure forming the joint support.

Here, we investigate the possibility to perform both US and PA super-resolution imaging of \emph{sparse samples} with model-based reconstruction approach using a \emph{sparse array}.
More specifically, we apply sparsity-constrained model-based reconstruction to perform 2D super-resolution imaging of sparse test samples with only 8 (out of 128) elements of a linear transducer array. We show that this reconstruction approach can be used for both PA and plane-wave US imaging with the same experimental setup. To build the matrix $\bs{A}$ that describes the forward model, we propose a novel method that involves only one PSF acquisition, as opposed to the measurement of the full set of PSFs in the imaging zone. We  also report simulation results showing how the reconstruction quality is related to the number of transducer elements and the signal-to-noise ratio (SNR).

\section*{Results}
\begin{figure}[!h]
\centering
\includegraphics[width=\linewidth]{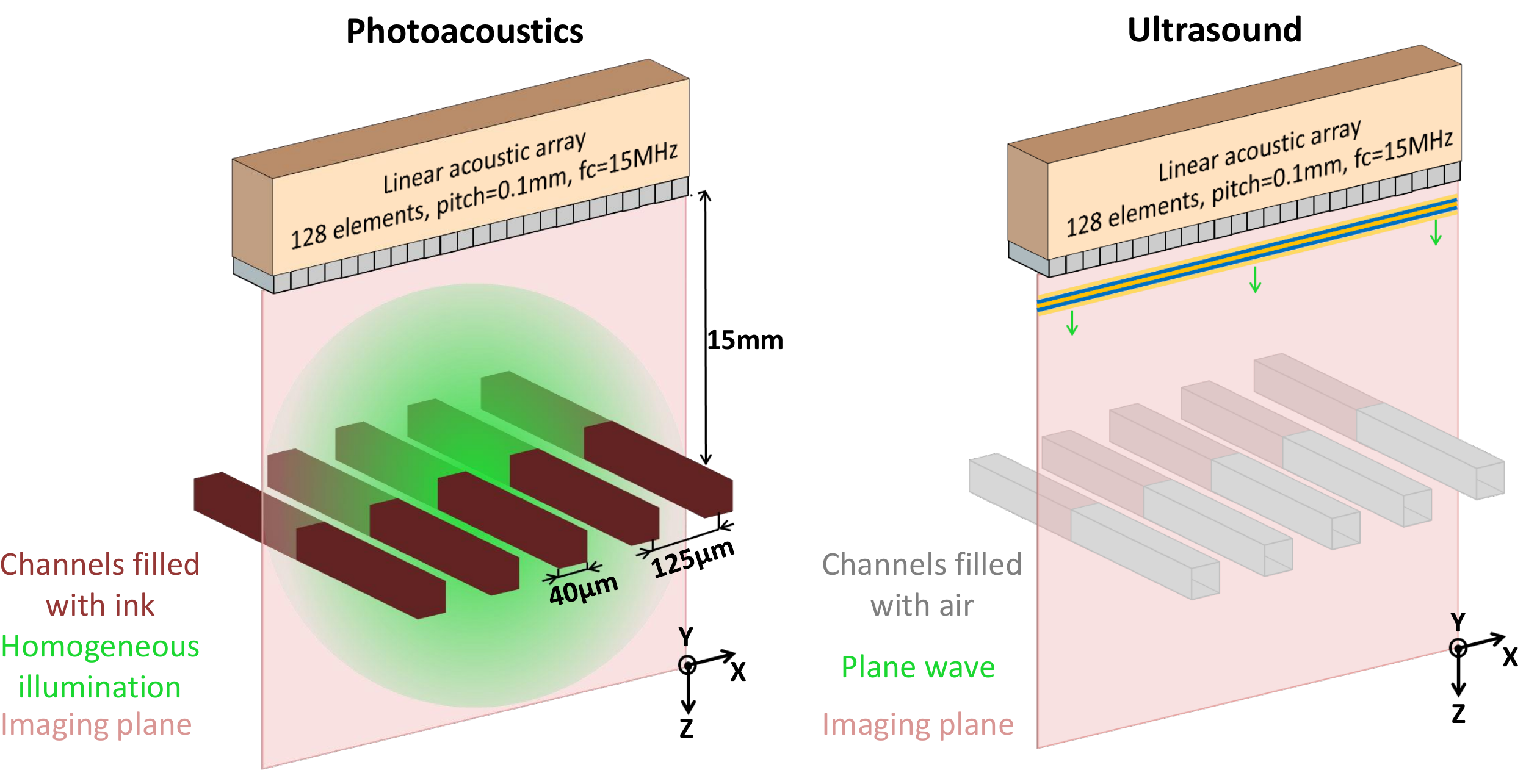}
\caption{Experimental setup. Samples carrying 5 parallel microchannels are placed perpendicularly to the imaging plane. Each channel is 40 $\mu$m wide and 50 $\mu$m deep, the center-to-center interchannel distance being $L_{cc}$ = 125 $\mu$m. In the PA experiment, the sample is illuminated by a laser pulse and the resulting PA signals are detected by a linear US probe. In the US experiment, a plane wave is emitted in the sample's direction and the backscattered signals are collected.}
\label{fig:experiments}
\end{figure}

To demonstrate that sparsity-constrained model-based reconstruction can provide super-resolution in sparse-array PA or US imaging, we carried out two proof-of-concept experiments (one for US and one for PA imaging). The goal of each experiment was to recover a super-resolved cross-sectional image of a sparse five-channel microfluidic sample by processing data received by different subsets of elements of a linear ultrasound probe. The imaging configuration for both experiments is shown in Fig. \ref{fig:experiments}. Importantly, in both PA and US experiments, we used the same imaging equipment, the same acquisition geometry, samples of identical structure, and the same reconstruction method. Only the nature of the contrast and the way the ultrasound wave is generated were different in these  experiments. In the PA experiment, the microfluidic channels were filled with absorbing liquid and illuminated by pulsed light. In the US experiment, the microfluidic channels were filled with air and a plane ultrasound wave was emitted by the transducer array. In both experiments, the measurement data consisted of the ultrasound signals detected by the array for a single shot excitation (light pulse or plane wave). 
Prior to imaging the five-channel samples, the PSF required to build the forward model was measured for each type of sample (absorber-filled or air-filled) in the corresponding imaging mode (PA or US). 

The image reconstruction consisted in finding the object that minimizes a cost function. This cost function was based on the forward model (derived from the PSF calibration) and a L1-norm regularization term used to suppress the measurement noise and select a sparse object. This model-based reconstruction  was then compared to the conventional delay-and-sum approach taken as a typical diffraction-limited  reconstruction approach. Further details on the experimental and reconstruction methods can be found in the Methods section. 

\begin{figure}[h]
\centering
\includegraphics[width=\linewidth]{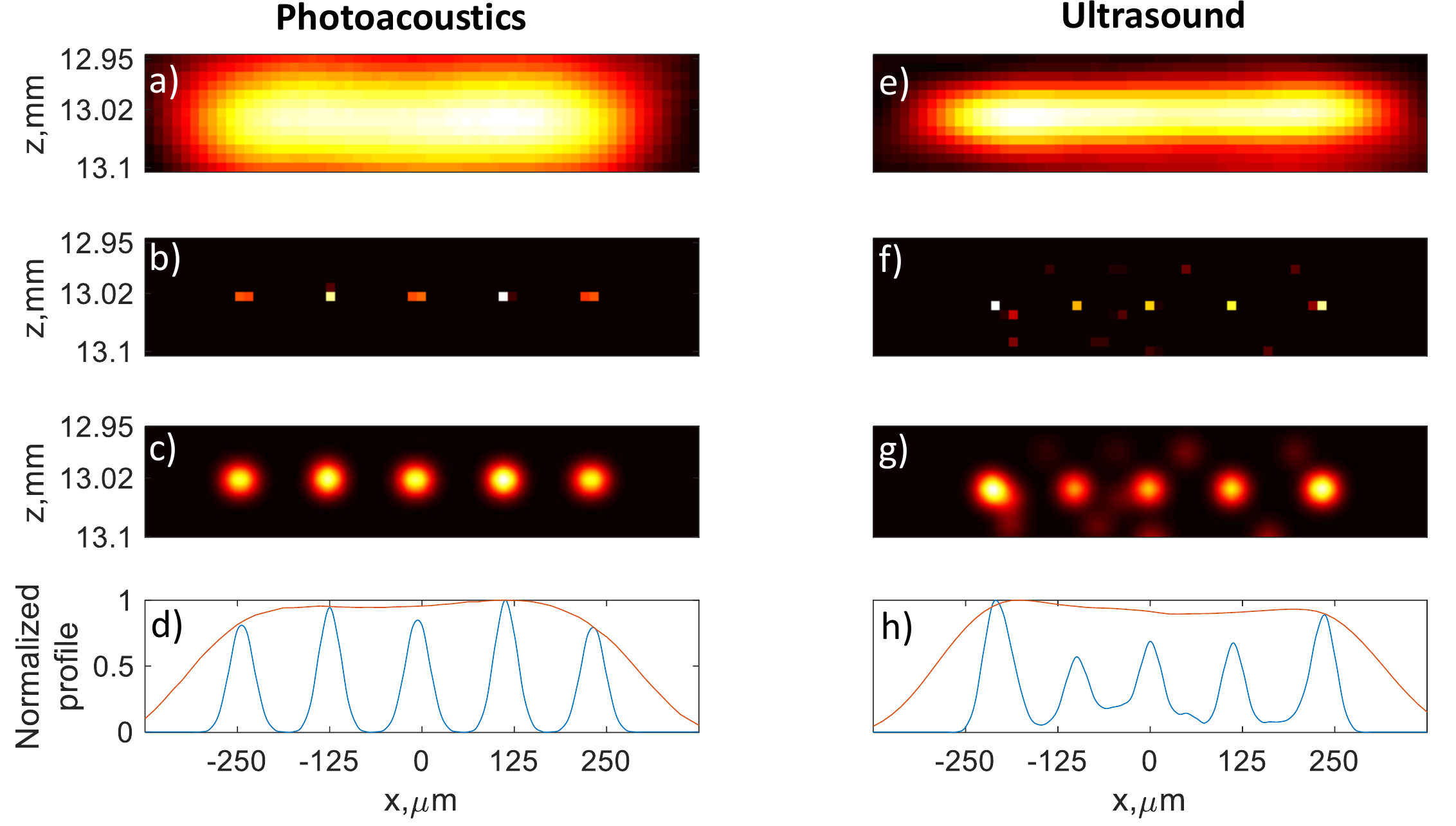}
\caption{Experimental images reconstructed in the PA (a-d) and US (e-h) experiments using all the 128 elements of the probe. (a,e) Conventional beamforming cross-sectional images of the sample. The microchannels
are indistinguishable as the center-to-center distance $L_{cc}$ = 125 $\mu$m is smaller than the lateral FWHM of the PSF (155 $\mu$m).
(b,f) Images obtained with model-based reconstruction. 
The reconstruction recovers five distinct regions corresponding to the microfluidic channels. 
(c,g) Reconstruction images (b) and (f) after smoothing out by a 2D spatial Gaussian filter ($\sigma$ = 12.5 $\mu$m) and interpolating on a 3.125 $\mu$m grid. (d,h) Red line: normalized amplitude profile on the envelope images (a) and (e); blue line: normalized amplitude profile on the filtered reconstruction images (c) and (g). }
\label{fig:res_128}
\end{figure}

We first demonstrated for both PA and US imaging that sparsity-based reconstruction led to super-resolution images of sparse samples when using all the 128 available elements of the probe. The conventional delay-and-sum images are shown in Fig. \ref{fig:res_128} a and e. It can be noticed that the conventional reconstruction is affected by diffraction-limited resolution: two neighboring channels can not be separated, resulting in a bar-like spot in place of the five individual channels. The center-to-center distance between neighbouring channels ($L_{cc}$ = 125 $\mu$m) is indeed below the diffraction limit defined by the lateral full width of half maximum (FWHM) of the PSF (measured to be 155 $\mu$m). Meanwhile, the five individual channels are clearly resolved in both PA and US sparsity-constrained model-based reconstruction images (Fig. \ref{fig:res_128} b-d and f-h). 

\begin{figure}[h]
\centering
\includegraphics[width=\linewidth]{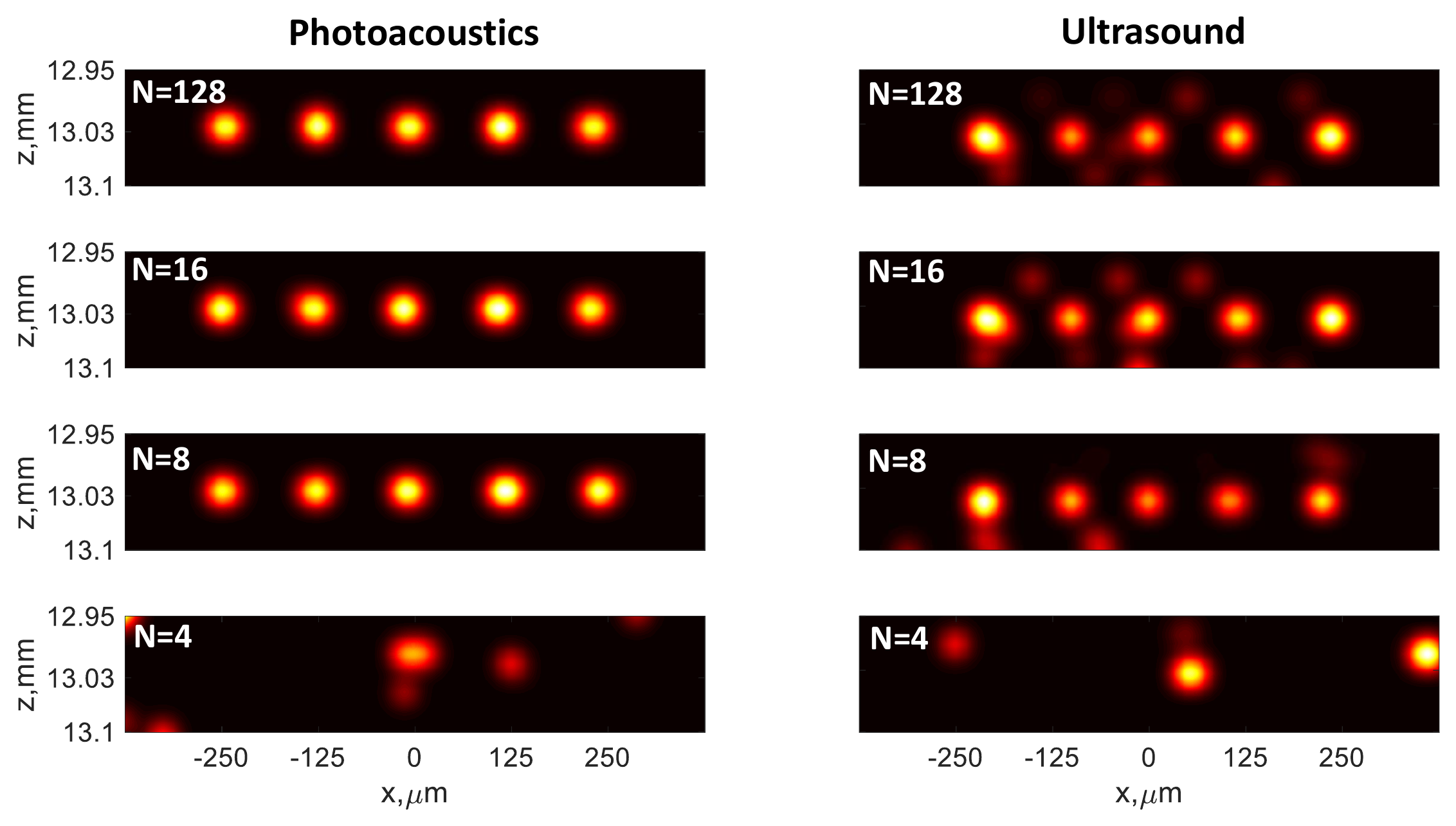}
\caption{Images obtained with model-based reconstruction in PA and US experiments using $N$ = 4, 8, 16 and 128 transducer elements regularly distributed along the probe aperture. A 2D spatial Gaussian filter ($\sigma$ = 12.5 $\mu$m) and interpolation on a 3.125 $\mu$m grid were applied after the reconstruction.}
\label{fig:sm_n_el}
\end{figure}

Then, as a major result of this work, we demonstrated that it could also be possible to obtain similar super-resolved images using a sparse array in the place of the initial 128-element dense array. To emulate sparse-array imaging, we applied the same reconstruction approach to data acquired by only a fraction of the probe's elements. To maintain the same conventional diffraction limit, the resulting  probe aperture was kept constant by including the first and the last elements of the probe. The other elements were regularly distributed along the probe. The objects reconstructed with 16, 8 and 4 elements are compared to the full-probe (128 elements) reconstruction in Fig. \ref{fig:sm_n_el}. 
The images in Fig. \ref{fig:sm_n_el} show that in both PA and US experiments the five channels could have been imaged using down to only 8 array elements.
With less than $N_{min}$ = 8 elements, we were not able to reconstruct the imaged structure properly. Random distributions of the transducer elements were also considered, but no significant differences in the results were found as compared to the linearly distributed case.

\begin{figure}[h]
\centering
\includegraphics[width=\linewidth]{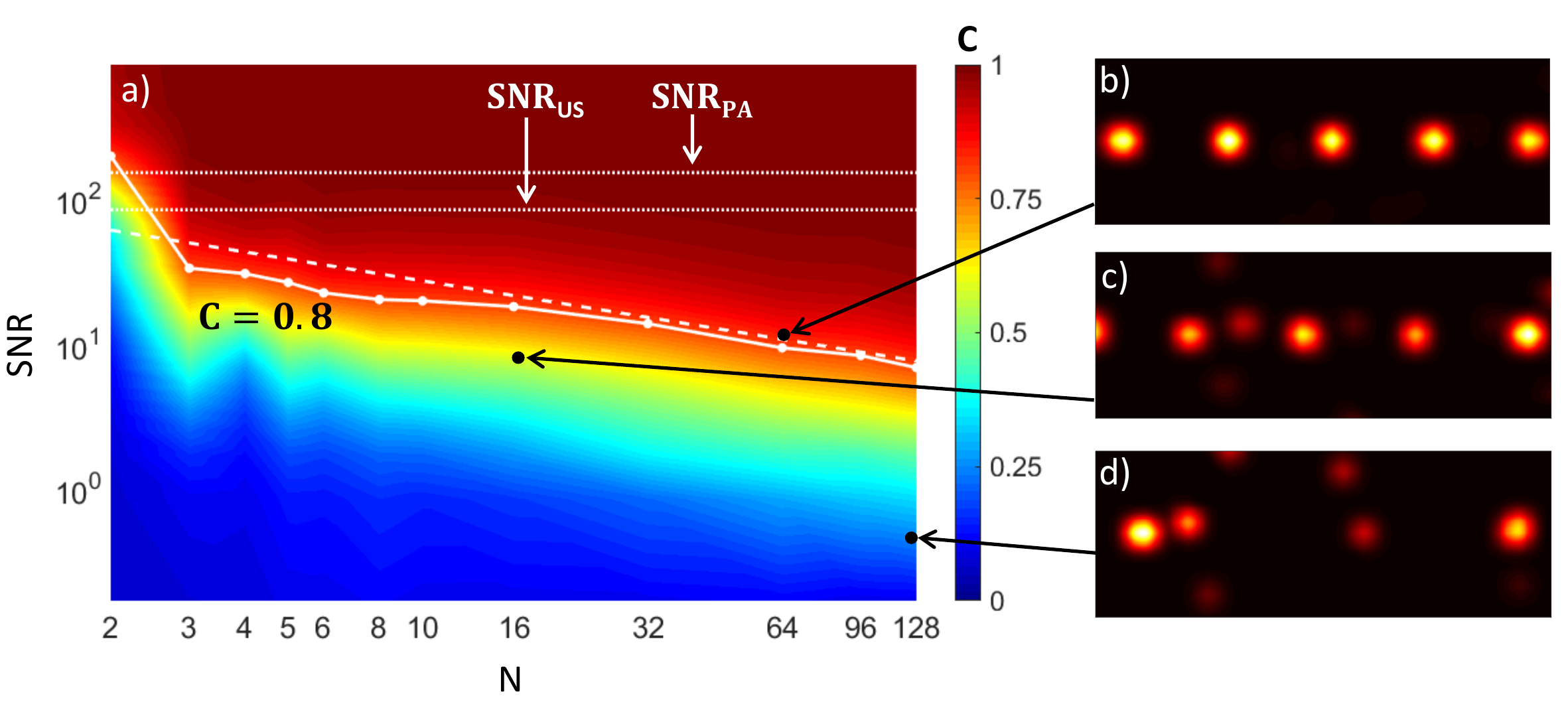}
\caption{Simulation results. (a) - Correlation $C$ as a function of the SNR and the number of transducer elements $N$, dotted lines: $SNR_{PA}$ = 150 in the PA experiment, $SNR_{US}$ = 83 in the US experiment, dashed line: function SNR$\propto \sqrt{N}$,  (b-d) - typical simulation images: (b) - $N$ = 64, SNR = 16, $C$ = 0.85, (c) - $N$ = 16, SNR = 10, $C$ = 0.57, (d) - $N$ = 128, SNR = 0.8, $C$ = 0.23.}
\label{fig:simulations}
\end{figure}

In order to shed some light on the parameters that condition the minimal number $N_{min}$ of transducer elements needed to yield a faithfull reconstruction,  we carried out a series of numerical simulations. In these simulations, as a measure of the reconstruction quality, we studied the correlation between the reconstructed object and a modelled ideal object, as a function of the number of transducer elements and the signal-to-noise ratio (SNR). In this work, we define the SNR as the ratio between the peak amplitude of the radio frequency (RF) signal and the standard deviation of the measurement noise computed over a signal-free region of the RF data. The simulation results are illustrated in Fig. \ref{fig:simulations}. As could be intuitively expected, these results show that for any fixed number $N$ of transducer elements involved in the reconstruction, the reconstruction image quality increases with the SNR. Moreover, the SNR that assures a given reconstruction quality approximately scales as the square root of the number of elements, apart from $N$ = 2 (the dashed line in Fig. \ref{fig:simulations} being traced for SNR$\propto \sqrt{N}$). For $N$ = 2, our simulations still predict the possibility to reconstruct the imaged object, but with a much stronger requirement on the SNR.

\section*{Discussion}

Previous studies reported super-resolution images obtained via model-based reconstruction with sparsity priors on the imaged object, in the field of US imaging~\cite{tuysuzoglu2012sparsity,zhao2016single} and more recently in PA imaging~\cite{egolf2018sparsity}. In this work, we first illustrated the generality of the sparsity-constrained model-based approach by applying this method to obtain experimental PA and US images of sparse test objects, with the same imaging equipment being used in the PA and US experiments. As a key results of the present work, we demonstrated that imaging sparse objects beyond the acoustic diffraction limit remains possible even when a very small number of transducer elements are used for the reconstruction, much below the number required by Nyquist spatial sampling.  This suggests that in imaging with conventional number of transducer elements, a significant part of the measurement information might be redundant. By means of numerical simulations, we showed that if the SNR is very high then by reconstructing images with a perfectly known forward model it is possible to obtain the correct object reconstruction with only two transducer elements. According to the simulation results,  more than two elements might be necessary to provide the correct reconstruction in the case of typical experimental values of the SNR: for a given SNR on each transducer element, using more elements may be considered  equivalent to reducing the influence of the measurement noise. In practice, additional uncertainties of the forward model might further limit the reconstruction quality: in our experiments the minimal number of elements $N$ = 8 required for the reconstruction with correlation $C$ = 0.8 was above the number of elements $N$ = 3 predicted by the simulations for the same SNR and the same correlation.

A certain difference in image reconstruction quality between PA images (left columns in Fig. \ref{fig:res_128} and \ref{fig:sm_n_el}) and US images (right columns in Fig. \ref{fig:res_128} and \ref{fig:sm_n_el}) can be noticed.  In particular, the SNR in the US reconstruction images is lower than in the PA images. In addition,  the average  interchannel distance in the US images (112 $\mu m$) slightly deviates from the true value (125 $\mu m$), whereas in the PA images the interchannel distance is restored correctly. This difference between the PA and US reconstruction images may result from several reasons. First, the SNR in the US experiment was lower than in the PA experiment. Second, the uncertainties of the forward model might have played a greater role in the US experiment.
As a novelty of our work, we derived the forward model by measuring only one PSF, rather than all PSFs in the field of view, which was done, for instance, in~\cite{egolf2018sparsity}. This  approach assumes that PSFs differ from each other only by time delays that are calculated based on wave propagation in water. However, in our experiments this condition was not fully satisfied at least due to the presence of a PDMS layer between the sources and the receivers. As the US experiment involves a double crossing of this PDMS layer, it is likely that in the US experiment the forward model is less accurate than in the PA experiment, leading to a less faithfull reconstruction image.   

In conclusion, we demonstrated experimentally the possibility of performing both PA and US super-resolution imaging of sparse samples with a sparse array (with down to 8 transducer elements) by applying the model-based reconstruction approach. Although the results demonstrated here were obtained in 2D imaging with a 1D linear array, a major advantage of the proposed approach is that it can be applied to any transducer geometry, and therefore in 3D imaging. Provided that the relative positions of the transducer elements are known, the proposed method of constructing the forward model from a single PSF measurement remains valid. 
It was shown by reconstructing  simulated data that the SNR yields a fundamental limit on the number of transducer elements needed to provide a faithfull reconstruction. The relative influence of the SNR and the uncertainties of the forward model on the reconstruction quality as well as the ultimate resolution limit are to be further investigated in order to fully understand and probably predict the number of transducer elements required for super-resolution imaging. 
While we demonstrated that it is possible to obtain super-resolution images of sparse test samples with sparse array, the presented method should be further investigated with more realistic experimental data before firm conclusions can be drawn on its performance in the context of biomedical PA and US imaging.

\section*{Methods}
\subsection*{Samples}
The microfluidic samples were prepared  with a standard soft-lithography manufacturing technology\cite{tang2009basic} using polydimethylsiloxane (PDMS). Each sample  consisted of five hollow channels sandwiched between two layers of PDMS. The thickness of the upper layer (the layer placed closer to the US probe in the experiments) was around 180 $\mu m$. This thickness was chosen as small as possible since the influence of the PDMS on wave propagation was neglected in the model. 
Each channel was 40 $\mu m$ wide (x-direction) and 50 $\mu m$ deep (z-direction). The centre-to-centre distance between neighbouring channels was $L_{cc}$ = 125 $\mu m$. This distance was deliberately chosen smaller than the estimated lateral full width of half maximum (FWHM) of the PSF (155 $\mu m$). To acquire the PSF and build the forward model, additional samples containing only one microfluidic channel were prepared, the only channel being 10 $\mu m$ wide and 50 $\mu m$ deep. For PA imaging, the channels were filled with an absorbing dye solution (Patent Blue V) to provide photoacoustic contrast, while for US imaging the channels were filled with air to provide acoustic contrast.

\subsection*{Experimental protocol}
The imaging configuration used in experiments is illustrated in Fig. \ref{fig:experiments}. The five microfluidic channels were aligned to cross the xz imaging plane perpendicularly, and positioned at the distance  $z_f$ = 15 mm from the linear ultrasound probe, the distance $z_f$ corresponding to the elevational focus of the probe.  As the US probe we used a capacitive micromachined ultrasonic (CMUT) array (L22-8v, Verasonics, USA: $N$ = 256 elements with 128 consecutive elements used in our experiments, pitch $\approx$ 100 $\mu m$, center frequency $f_c\approx$ 15 MHz). 
The probe and sample were immersed in a water tank. In the PA experiment, signals were generated by illuminating the sample with a 5 ns laser pulse ($\lambda$ = 532 nm, fluence $\approx$ 3 mJ/cm$^2$) generated by a frequency-doubled Nd:YAG laser (Spitlight DPSS 250, Innolas Laser GmbH, Krailling, Germany).  In the US experiment, a plane ultrasound wave (15 MHz center frequency short pulse) was emitted by the probe, and ultrasound waves backscattered from the sample were recorded. To control emission and reception of ultrasound waves, the probe was connected to multichannel acquisition electronics (High Frequency Vantage 256, Verasonics, USA). 

To reconstruct objects in both PA and US experiments, the PSF was measured by applying the acquisition protocol described above for samples containing a single microfluidic channel. The only channel was placed in the center of the field of view of the ultrasound probe.  The conventional lateral resolution defined as the lateral full width at half maximum (FWHM) of the PSF was measured after standard delay-and-sum reconstruction and turned out to be around 155 $\mu$m in both PA and US experiments.

\subsection*{Forward model}

Under the assumption of a linear dependence between the quantity to image and the measured radio frequency (RF) signals, the voltage signals measured by the acquisition electronics can be expressed for both PA and US imaging as  
\begin{equation}
    s(t_i,\rr_k)=\mathbb{A}[T_0(\rr)],
    \label{exp:sparsity_intro_H_combined_short}
\end{equation}
where $s(t_i,\rr_k)$ is the signal at time $t_k$ measured by the transducer element located at $\rr_k$, $T_0(\rr)$ is the quantity to image, and $\mathbb{A}$ is a linear operator that takes into account both the ultrasound generation/propagation, and the transducer response. In PA imaging, $T_0(\rr)$ is proportional to the absorption coefficient $\mu_a(\rr)$, under the assumption of homogeneous light illumination and a homogeneous Gruneisen paramater. In US imaging,  $T_0(\rr)$ is related to the distribution of the backscattering coefficient, provided that the single scattering regime is valid.

By discretizing the object to reconstruct on a grid, this forward problem may then be written in a matrix form as 
\begin{equation}
S=\bs{A}T_0,
\label{exp:forward_model}
\end{equation}
where $S^{m \times 1}$ is a vector with all the RF data (m =  number of time samples $\times$  number of transducer elements), $T_0^{n \times 1}$ is the discretized version of the quantity to reconstruct (n = number of grid points in the reconstruction zone), and $\bs{A}^{m \times n}$ is a matrix representing the linear operator.  
Each of the $n$ columns of $\bs{A}$ represents the RF response for one of the points $\overline{1..n}$ of the reconstruction zone, i.e. each column of $\bs{A}$ is the vector data corresponding to the signals from a single point source in the imaging zone.  The matrix $\bs{A}$ therefore  contains  the responses to each point source and can be considered as a matrix of all the system point spread functions (PSF) in the reconstruction space. 

The matrix $\bs{A}$ can be modelled theoretically or measured experimentally. In some previous works, $\bs{A}$ was measured for each point of the reconstruction zone\cite{egolf2018sparsity,roumeliotis2011singular,roumeliotis20123d, murray2017super}. In our work, the matrix $\bs{A}$ is also obtained experimentally but we perform a measurement for only one point source. As a point source, we use a single isolated microfluidic channel.
All the columns of $\bs{A}$ are then derived from this measurement data assuming that although the signals from two distinct point sources  differ from each other by their arrival time, apart from this they are identical. In other words, it is assumed that the response of each transducer element is the same for all  points in the reconstruction zone.  In our experiments, we also neglect the presence of the upper PDMS layer and therefore only consider wave propagation in water, i.e. in a homogeneous isotropic medium with a constant speed of sound $c$ = 1500 $m/s$.

Under the assumptions stated above, when a single PSF is acquired for a source placed at $\{x_j,z_j\}$, any column $i$ of the matrix $\bs{A}$ can be derived from the acquired RF data by shifting the signals for each transducer element $k$ with the following time delay : 
\begin{equation}
\Delta t_{i,j,k}=\delta_{US}\frac{z_j-z_i}{c}+\frac{1}{c}\Big(\sqrt{(x_k-x_i)^2+(z_k-z_i)^2}-\sqrt{(x_k-x_j)^2+(z_k-z_j)^2}\Big),
\label{exp:travel_time}
\end{equation}
where $\{x_k,z_k\}$ are the coordinates of element $k$ of the transducer array, $\{x_i,z_i\}$ are the coordinates of point $i$ of the reconstruction grid. In US imaging, $\delta_{US}$ = 1  to account for the travel time of the emitted plane wave ($\delta_{US}$ = 0 in PA imaging).

\subsection*{Image reconstruction}

To solve the inverse problem, i.e. to find an estimate $\hat{T}$ of the solution $T_0$ to the forward model described above, the sparsity-constrained minimization approach similar to those already reported in previous works \cite{egolf2018sparsity,tuysuzoglu2012sparsity,murray2017super} was used. This approach consists in finding $\hat{T}$ by solving  the following minimization problem: 
\begin{equation}
\hat{T}=\underset{T}{\text{argmin}}\{||S-\bs{A}T||^2_2+\alpha^2||T||_1\}.
\label{exp:regularization}
\end{equation}
In Eq. (\ref{exp:regularization}) the $l_2$ - term $||S-\bs{A}T||^2_2=\sum_{\text{all data points } p}(S-\bs{A}T)^2_p$ corresponds to the least square fitting of the probed quantity  $T$ to the acquired data $S$. The regularization $l_1$ - term $\alpha^2||T||_1=\alpha^2\sum_{p}|T_p|$ is used to minimize the influence of the noise in $S$ while selecting a sparse solution to underdetermined system (\ref{exp:forward_model}). Regarding typical scales involved in the reconstruction, for reconstruction with 128 transducer elements the length of the vector $T^{n\times 1}$ was $n$ = 793 (number of reconstruction points in the field of view), the length of the vector $S^{m\times 1}$ was $m$ = 4636, the rank of the matrix $\bs{A}$ was $\text{rank}(A)$ = 346. The rank of the matrix $\bs{A}$ being inferior to the number of unknowns $n$, system (\ref{exp:forward_model}) was undetermined.
The minimization operation described by Eq.(\ref{exp:regularization}) was performed using a fast iterative shrinkage-thresholding algorithm (FISTA)\cite{beck2009fast, palomar2010convex, FISTAMATLAB}. 
The object to image was reconstructed over a cartesian grid with a $L_{cc}$/10 = 12.5 $\mu$m step. As in previous works\cite{tuysuzoglu2012sparsity,murray2017super,egolf2018sparsity}, the value of the regularization parameter $\alpha$ was determined heuristically such that the reconstruction image was qualitatively close to the real sample. A 2D Gaussian filter and interpolation were applied to improve image visualization. The final image definition corresponds to a grid step of $3.125 \mu m$.

\subsection*{Simulations}
Numerical simulations were performed to produce test data corresponding to imaging five point sources.  The stated  assumptions on our forward model were strictly followed. Specifically, we modelled the signals received on each transducer element as a time-delayed version of the same signal (simulated for five point sources), based on the delay law described by Eq.~\ref{exp:travel_time}.
The detected signals had a central frequency and bandwidth corresponding to those used in experiments and sampled at the same  frequency as in experiments. Simulation data with different SNR values was generated  by varying the amplitude of the modelled PA signals. In all simulations, Gaussian noise with a zero mean and the rms of $\sigma_n=30$ was added to the detected signals. Such noise corresponded to the noise produced by the acquisition electronics in our experiments.
The simulated distribution of sources was then reconstructed following the same methods as used to reconstruct images from the experimental data. As a metrics of the reconstruction quality, we computed a normalized spatial cross-correlation between each reconstructed object $T$ and the ideal reconstruction $T_{ref}$.

The ideal reconstruction $T_{ref}$ was modeled by asserting the value of 1 to the 5 cells of the reconstruction grid corresponding to the positions of the 5 simulated sources. Then, the filtering and interpolation used for the reconstruction images were also applied to $T_{ref}$.  The correlation value for each SNR was estimated from  100 noise realizations. Fig. \ref{fig:simulations} represents the average correlation for each SNR value.

\section*{Funding}

This project has received funding from the European Research Council (ERC) under the European Union’s Horizon 2020 research and innovation program (grants no. 278025, 677909, 681514), and Human Frontiers Science Program. O.K. was supported by an Azrieli Faculty Fellowship.

\bibliography{main}

\begin{thebibliography}{10}
\urlstyle{rm}
\expandafter\ifx\csname url\endcsname\relax
  \def\url#1{\texttt{#1}}\fi
\expandafter\ifx\csname urlprefix\endcsname\relax\def\urlprefix{URL }\fi
\expandafter\ifx\csname doiprefix\endcsname\relax\def\doiprefix{DOI: }\fi
\providecommand{\bibinfo}[2]{#2}
\providecommand{\eprint}[2][]{\url{#2}}

\bibitem{wells2006ultrasound}
\bibinfo{author}{Wells, P.~N.}
\newblock \bibinfo{journal}{\bibinfo{title}{Ultrasound imaging}}.
\newblock {\emph{\JournalTitle{Physics in Medicine \& Biology}}}
  \textbf{\bibinfo{volume}{51}}, \bibinfo{pages}{R83} (\bibinfo{year}{2006}).

\bibitem{beard2011biomedical}
\bibinfo{author}{Beard, P.}
\newblock \bibinfo{journal}{\bibinfo{title}{Biomedical photoacoustic imaging}}.
\newblock {\emph{\JournalTitle{Interface focus}}} \textbf{\bibinfo{volume}{1}},
  \bibinfo{pages}{602--631} (\bibinfo{year}{2011}).

\bibitem{Fenster_2001}
\bibinfo{author}{Fenster, A.}, \bibinfo{author}{Downey, D.~B.} \&
  \bibinfo{author}{Cardinal, H.~N.}
\newblock \bibinfo{journal}{\bibinfo{title}{Three-dimensional ultrasound
  imaging}}.
\newblock {\emph{\JournalTitle{Physics in Medicine and Biology}}}
  \textbf{\bibinfo{volume}{46}}, \bibinfo{pages}{R67--R99}
  (\bibinfo{year}{2001}).

\bibitem{turnbull1991beam}
\bibinfo{author}{Turnbull, D.~H.} \& \bibinfo{author}{Foster, F.~S.}
\newblock \bibinfo{journal}{\bibinfo{title}{Beam steering with pulsed
  two-dimensional transducer arrays}}.
\newblock {\emph{\JournalTitle{IEEE transactions on ultrasonics,
  ferroelectrics, and frequency control}}} \textbf{\bibinfo{volume}{38}},
  \bibinfo{pages}{320--333} (\bibinfo{year}{1991}).

\bibitem{smith1991high}
\bibinfo{author}{Smith, S.~W.}, \bibinfo{author}{Pavy, H.~G.} \&
  \bibinfo{author}{von Ramm, O.~T.}
\newblock \bibinfo{journal}{\bibinfo{title}{High-speed ultrasound volumetric
  imaging system. i. transducer design and beam steering}}.
\newblock {\emph{\JournalTitle{IEEE transactions on ultrasonics,
  ferroelectrics, and frequency control}}} \textbf{\bibinfo{volume}{38}},
  \bibinfo{pages}{100--108} (\bibinfo{year}{1991}).

\bibitem{roux20173d}
\bibinfo{author}{Roux, E.} \emph{et~al.}
\newblock \bibinfo{title}{3d diverging waves with 2d sparse arrays: A
  feasibility study}.
\newblock In \emph{\bibinfo{booktitle}{2017 IEEE International Ultrasonics
  Symposium (IUS)}}, \bibinfo{pages}{1--4} (\bibinfo{organization}{IEEE},
  \bibinfo{year}{2017}).

\bibitem{roux2017validation}
\bibinfo{author}{Roux, E.} \emph{et~al.}
\newblock \bibinfo{title}{Validation of optimal 2d sparse arrays in focused
  mode: Phantom experiments}.
\newblock In \emph{\bibinfo{booktitle}{2017 IEEE International Ultrasonics
  Symposium (IUS)}}, \bibinfo{pages}{1--4} (\bibinfo{organization}{IEEE},
  \bibinfo{year}{2017}).

\bibitem{cohen2018sparse}
\bibinfo{author}{Cohen, R.} \& \bibinfo{author}{Eldar, Y.~C.}
\newblock \bibinfo{journal}{\bibinfo{title}{Sparse convolutional beamforming
  for ultrasound imaging}}.
\newblock {\emph{\JournalTitle{IEEE transactions on ultrasonics,
  ferroelectrics, and frequency control}}} \textbf{\bibinfo{volume}{65}},
  \bibinfo{pages}{2390--2406} (\bibinfo{year}{2018}).

\bibitem{eldar2012compressed}
\bibinfo{author}{Eldar, Y.~C.} \& \bibinfo{author}{Kutyniok, G.}
\newblock \emph{\bibinfo{title}{Compressed sensing: theory and applications}}
  (\bibinfo{publisher}{Cambridge university press}, \bibinfo{year}{2012}).

\bibitem{rossi2013spatial}
\bibinfo{author}{Rossi, M.}, \bibinfo{author}{Haimovich, A.~M.} \&
  \bibinfo{author}{Eldar, Y.~C.}
\newblock \bibinfo{journal}{\bibinfo{title}{Spatial compressive sensing for
  mimo radar}}.
\newblock {\emph{\JournalTitle{IEEE Transactions on Signal Processing}}}
  \textbf{\bibinfo{volume}{62}}, \bibinfo{pages}{419--430}
  (\bibinfo{year}{2013}).

\bibitem{cohen2018summer}
\bibinfo{author}{Cohen, D.}, \bibinfo{author}{Cohen, D.},
  \bibinfo{author}{Eldar, Y.~C.} \& \bibinfo{author}{Haimovich, A.~M.}
\newblock \bibinfo{journal}{\bibinfo{title}{Summer: Sub-nyquist mimo radar}}.
\newblock {\emph{\JournalTitle{IEEE Transactions on Signal Processing}}}
  \textbf{\bibinfo{volume}{66}}, \bibinfo{pages}{4315--4330}
  (\bibinfo{year}{2018}).

\bibitem{ephrat2008three}
\bibinfo{author}{Ephrat, P.}, \bibinfo{author}{Keenlislide, L.},
  \bibinfo{author}{Seabrook, A.}, \bibinfo{author}{Prato, F.~S.} \&
  \bibinfo{author}{Carson, J.~J.}
\newblock \bibinfo{journal}{\bibinfo{title}{Three-dimensional photoacoustic
  imaging by sparse-array detection and iterative image reconstruction}}.
\newblock {\emph{\JournalTitle{Journal of Biomedical Optics}}}
  \textbf{\bibinfo{volume}{13}}, \bibinfo{pages}{054052}
  (\bibinfo{year}{2008}).

\bibitem{chaudhary2010comparison}
\bibinfo{author}{Chaudhary, G.}, \bibinfo{author}{Roumeliotis, M.},
  \bibinfo{author}{Carson, J.} \& \bibinfo{author}{Anastasio, M.}
\newblock \bibinfo{title}{Comparison of reconstruction algorithms for
  sparse-array detection photoacoustic tomography}.
\newblock In \emph{\bibinfo{booktitle}{Photons Plus Ultrasound: Imaging and
  Sensing 2010}}, vol. \bibinfo{volume}{7564}, \bibinfo{pages}{756434}
  (\bibinfo{organization}{International Society for Optics and Photonics},
  \bibinfo{year}{2010}).

\bibitem{roumeliotis2011singular}
\bibinfo{author}{Roumeliotis, M.~B.}, \bibinfo{author}{Stodilka, R.~Z.},
  \bibinfo{author}{Anastasio, M.~A.}, \bibinfo{author}{Ng, E.} \&
  \bibinfo{author}{Carson, J.~J.}
\newblock \bibinfo{journal}{\bibinfo{title}{Singular value decomposition
  analysis of a photoacoustic imaging system and 3d imaging at 0.7 fps}}.
\newblock {\emph{\JournalTitle{Optics express}}} \textbf{\bibinfo{volume}{19}},
  \bibinfo{pages}{13405--13417} (\bibinfo{year}{2011}).

\bibitem{roumeliotis20123d}
\bibinfo{author}{Roumeliotis, M.~B.}, \bibinfo{author}{Kosik, I.} \&
  \bibinfo{author}{Carson, J.~J.}
\newblock \bibinfo{title}{3d photoacoustic imaging using a staring-sparse array
  with 60 transducers}.
\newblock In \emph{\bibinfo{booktitle}{Photons Plus Ultrasound: Imaging and
  Sensing 2012}}, vol. \bibinfo{volume}{8223}, \bibinfo{pages}{82233F}
  (\bibinfo{organization}{International Society for Optics and Photonics},
  \bibinfo{year}{2012}).

\bibitem{wang2012investigation}
\bibinfo{author}{Wang, K.}, \bibinfo{author}{Su, R.},
  \bibinfo{author}{Oraevsky, A.~A.} \& \bibinfo{author}{Anastasio, M.~A.}
\newblock \bibinfo{journal}{\bibinfo{title}{Investigation of iterative image
  reconstruction in three-dimensional optoacoustic tomography}}.
\newblock {\emph{\JournalTitle{Physics in Medicine \& Biology}}}
  \textbf{\bibinfo{volume}{57}}, \bibinfo{pages}{5399} (\bibinfo{year}{2012}).

\bibitem{dean2012accurate}
\bibinfo{author}{Dean-Ben, X.~L.}, \bibinfo{author}{Buehler, A.},
  \bibinfo{author}{Ntziachristos, V.} \& \bibinfo{author}{Razansky, D.}
\newblock \bibinfo{journal}{\bibinfo{title}{Accurate model-based reconstruction
  algorithm for three-dimensional optoacoustic tomography}}.
\newblock {\emph{\JournalTitle{IEEE Transactions on Medical Imaging}}}
  \textbf{\bibinfo{volume}{31}}, \bibinfo{pages}{1922--1928}
  (\bibinfo{year}{2012}).

\bibitem{han2017three}
\bibinfo{author}{Han, Y.} \emph{et~al.}
\newblock \bibinfo{journal}{\bibinfo{title}{Three-dimensional optoacoustic
  reconstruction using fast sparse representation}}.
\newblock {\emph{\JournalTitle{Optics letters}}} \textbf{\bibinfo{volume}{42}},
  \bibinfo{pages}{979--982} (\bibinfo{year}{2017}).

\bibitem{foster2000advances}
\bibinfo{author}{Foster, F.~S.}, \bibinfo{author}{Pavlin, C.~J.},
  \bibinfo{author}{Harasiewicz, K.~A.}, \bibinfo{author}{Christopher, D.~A.} \&
  \bibinfo{author}{Turnbull, D.~H.}
\newblock \bibinfo{journal}{\bibinfo{title}{Advances in ultrasound
  biomicroscopy}}.
\newblock {\emph{\JournalTitle{Ultrasound in medicine \& biology}}}
  \textbf{\bibinfo{volume}{26}}, \bibinfo{pages}{1--27} (\bibinfo{year}{2000}).

\bibitem{betzig2006imaging}
\bibinfo{author}{Betzig, E.} \emph{et~al.}
\newblock \bibinfo{journal}{\bibinfo{title}{Imaging intracellular fluorescent
  proteins at nanometer resolution}}.
\newblock {\emph{\JournalTitle{Science}}} \textbf{\bibinfo{volume}{313}},
  \bibinfo{pages}{1642--1645} (\bibinfo{year}{2006}).

\bibitem{hell1994breaking}
\bibinfo{author}{Hell, S.~W.} \& \bibinfo{author}{Wichmann, J.}
\newblock \bibinfo{journal}{\bibinfo{title}{Breaking the diffraction resolution
  limit by stimulated emission: stimulated-emission-depletion fluorescence
  microscopy}}.
\newblock {\emph{\JournalTitle{Optics letters}}} \textbf{\bibinfo{volume}{19}},
  \bibinfo{pages}{780--782} (\bibinfo{year}{1994}).

\bibitem{viessmann2013acoustic}
\bibinfo{author}{Viessmann, O.}, \bibinfo{author}{Eckersley, R.},
  \bibinfo{author}{Christensen-Jeffries, K.}, \bibinfo{author}{Tang, M.} \&
  \bibinfo{author}{Dunsby, C.}
\newblock \bibinfo{journal}{\bibinfo{title}{Acoustic super-resolution with
  ultrasound and microbubbles}}.
\newblock {\emph{\JournalTitle{Physics in Medicine \& Biology}}}
  \textbf{\bibinfo{volume}{58}}, \bibinfo{pages}{6447} (\bibinfo{year}{2013}).

\bibitem{oreilly2013super}
\bibinfo{author}{O'Reilly, M.~A.} \& \bibinfo{author}{Hynynen, K.}
\newblock \bibinfo{journal}{\bibinfo{title}{A super-resolution ultrasound
  method for brain vascular mapping}}.
\newblock {\emph{\JournalTitle{Medical physics}}} \textbf{\bibinfo{volume}{40}}
  (\bibinfo{year}{2013}).

\bibitem{desailly2013sono}
\bibinfo{author}{Desailly, Y.}, \bibinfo{author}{Couture, O.},
  \bibinfo{author}{Fink, M.} \& \bibinfo{author}{Tanter, M.}
\newblock \bibinfo{journal}{\bibinfo{title}{Sono-activated ultrasound
  localization microscopy}}.
\newblock {\emph{\JournalTitle{Applied Physics Letters}}}
  \textbf{\bibinfo{volume}{103}}, \bibinfo{pages}{174107}
  (\bibinfo{year}{2013}).

\bibitem{vilov2017overcoming}
\bibinfo{author}{Vilov, S.}, \bibinfo{author}{Arnal, B.} \&
  \bibinfo{author}{Bossy, E.}
\newblock \bibinfo{journal}{\bibinfo{title}{Overcoming the acoustic diffraction
  limit in photoacoustic imaging by the localization of flowing absorbers}}.
\newblock {\emph{\JournalTitle{Optics letters}}} \textbf{\bibinfo{volume}{42}},
  \bibinfo{pages}{4379--4382} (\bibinfo{year}{2017}).

\bibitem{dean2018localization}
\bibinfo{author}{Dean-Ben, X.~L.} \& \bibinfo{author}{Razansky, D.}
\newblock \bibinfo{journal}{\bibinfo{title}{Localization optoacoustic
  tomography}}.
\newblock {\emph{\JournalTitle{Light: Science \& Applications}}}
  \textbf{\bibinfo{volume}{7}}, \bibinfo{pages}{18004} (\bibinfo{year}{2018}).

\bibitem{chaigne2016super}
\bibinfo{author}{Chaigne, T.} \emph{et~al.}
\newblock \bibinfo{journal}{\bibinfo{title}{Super-resolution photoacoustic
  fluctuation imaging with multiple speckle illumination}}.
\newblock {\emph{\JournalTitle{Optica}}} \textbf{\bibinfo{volume}{3}},
  \bibinfo{pages}{54--57} (\bibinfo{year}{2016}).

\bibitem{chaigne2017super}
\bibinfo{author}{Chaigne, T.}, \bibinfo{author}{Arnal, B.},
  \bibinfo{author}{Vilov, S.}, \bibinfo{author}{Bossy, E.} \&
  \bibinfo{author}{Katz, O.}
\newblock \bibinfo{journal}{\bibinfo{title}{Super-resolution photoacoustic
  imaging via flow-induced absorption fluctuations}}.
\newblock {\emph{\JournalTitle{Optica}}} \textbf{\bibinfo{volume}{4}},
  \bibinfo{pages}{1397--1404} (\bibinfo{year}{2017}).

\bibitem{bar2017}
\bibinfo{author}{{Bar-Zion}, A.}, \bibinfo{author}{{Tremblay-Darveau}, C.},
  \bibinfo{author}{{Solomon}, O.}, \bibinfo{author}{{Adam}, D.} \&
  \bibinfo{author}{{Eldar}, Y.~C.}
\newblock \bibinfo{journal}{\bibinfo{title}{Fast vascular ultrasound imaging
  with enhanced spatial resolution and background rejection}}.
\newblock {\emph{\JournalTitle{IEEE Transactions on Medical Imaging}}}
  \textbf{\bibinfo{volume}{36}}, \bibinfo{pages}{169--180}
  (\bibinfo{year}{2017}).

\bibitem{tuysuzoglu2012sparsity}
\bibinfo{author}{Tuysuzoglu, A.}, \bibinfo{author}{Kracht, J.~M.},
  \bibinfo{author}{Cleveland, R.~O.}, \bibinfo{author}{C{\c{}}~etin, M.} \&
  \bibinfo{author}{Karl, W.~C.}
\newblock \bibinfo{journal}{\bibinfo{title}{Sparsity driven ultrasound
  imaging}}.
\newblock {\emph{\JournalTitle{The Journal of the Acoustical Society of
  America}}} \textbf{\bibinfo{volume}{131}}, \bibinfo{pages}{1271--1281}
  (\bibinfo{year}{2012}).

\bibitem{zhao2016single}
\bibinfo{author}{Zhao, N.}, \bibinfo{author}{Wei, Q.},
  \bibinfo{author}{Basarab, A.}, \bibinfo{author}{Kouam{\'e}, D.} \&
  \bibinfo{author}{Tourneret, J.-Y.}
\newblock \bibinfo{title}{Single image super-resolution of medical ultrasound
  images using a fast algorithm}.
\newblock In \emph{\bibinfo{booktitle}{Biomedical Imaging (ISBI), 2016 IEEE
  13th International Symposium on}}, \bibinfo{pages}{473--476}
  (\bibinfo{organization}{IEEE}, \bibinfo{year}{2016}).

\bibitem{egolf2018sparsity}
\bibinfo{author}{Egolf, D.~M.}, \bibinfo{author}{Chee, R.~K.} \&
  \bibinfo{author}{Zemp, R.~J.}
\newblock \bibinfo{journal}{\bibinfo{title}{Sparsity-based reconstruction for
  super-resolved limited-view photoacoustic computed tomography deep in a
  scattering medium}}.
\newblock {\emph{\JournalTitle{Optics letters}}} \textbf{\bibinfo{volume}{43}},
  \bibinfo{pages}{2221--2224} (\bibinfo{year}{2018}).

\bibitem{hojman2017photoacoustic}
\bibinfo{author}{Hojman, E.} \emph{et~al.}
\newblock \bibinfo{journal}{\bibinfo{title}{Photoacoustic imaging beyond the
  acoustic diffraction-limit with dynamic speckle illumination and sparse joint
  support recovery}}.
\newblock {\emph{\JournalTitle{Optics express}}} \textbf{\bibinfo{volume}{25}},
  \bibinfo{pages}{4875--4886} (\bibinfo{year}{2017}).

\bibitem{murray2017super}
\bibinfo{author}{Murray, T.~W.}, \bibinfo{author}{Haltmeier, M.},
  \bibinfo{author}{Berer, T.}, \bibinfo{author}{Leiss-Holzinger, E.} \&
  \bibinfo{author}{Burgholzer, P.}
\newblock \bibinfo{journal}{\bibinfo{title}{Super-resolution photoacoustic
  microscopy using blind structured illumination}}.
\newblock {\emph{\JournalTitle{Optica}}} \textbf{\bibinfo{volume}{4}},
  \bibinfo{pages}{17--22} (\bibinfo{year}{2017}).

\bibitem{bar2018sparsity}
\bibinfo{author}{{Bar-Zion}, A.}, \bibinfo{author}{{Solomon}, O.},
  \bibinfo{author}{{Tremblay-Darveau}, C.}, \bibinfo{author}{{Adam}, D.} \&
  \bibinfo{author}{{Eldar}, Y.~C.}
\newblock \bibinfo{journal}{\bibinfo{title}{Sushi: Sparsity-based ultrasound
  super-resolution hemodynamic imaging}}.
\newblock {\emph{\JournalTitle{IEEE Transactions on Ultrasonics,
  Ferroelectrics, and Frequency Control}}} \textbf{\bibinfo{volume}{65}},
  \bibinfo{pages}{2365--2380} (\bibinfo{year}{2018}).

\bibitem{tang2009basic}
\bibinfo{author}{Tang, S.~K.} \& \bibinfo{author}{Whitesides, G.~M.}
\newblock \bibinfo{journal}{\bibinfo{title}{Basic microfluidic and soft
  lithographic techniques}}.
\newblock {\emph{\JournalTitle{Optofluidics: Fundamentals, Devices and
  Applications, Y. Fainman, L. Lee, D. Psaltis, and C. Yang, eds (McGraw-Hill,
  2010)}}}  (\bibinfo{year}{2009}).

\bibitem{beck2009fast}
\bibinfo{author}{Beck, A.} \& \bibinfo{author}{Teboulle, M.}
\newblock \bibinfo{journal}{\bibinfo{title}{A fast iterative
  shrinkage-thresholding algorithm for linear inverse problems}}.
\newblock {\emph{\JournalTitle{SIAM journal on imaging sciences}}}
  \textbf{\bibinfo{volume}{2}}, \bibinfo{pages}{183--202}
  (\bibinfo{year}{2009}).

\bibitem{palomar2010convex}
\bibinfo{author}{Palomar, D.~P.} \& \bibinfo{author}{Eldar, Y.~C.}
\newblock \emph{\bibinfo{title}{Convex optimization in signal processing and
  communications}} (\bibinfo{publisher}{Cambridge university press},
  \bibinfo{year}{2010}).

\bibitem{FISTAMATLAB}
\bibinfo{author}{Vu, T.}
\newblock \bibinfo{title}{Fista implementation in matlab}.
\newblock \bibinfo{howpublished}{\url{https://github.com/tiepvupsu/FISTA}}
  (\bibinfo{year}{2016}).

\end{thebibliography}

\end{document}